\documentclass[aps,prb,twocolumn,superscriptaddress]{revtex4}
\usepackage{amsmath}
\usepackage{amssymb}
\usepackage{graphicx}

\usepackage{color}
\usepackage{bm}
\usepackage{epstopdf}

\begin{document}

\title{Bistability phenomena in 1D polariton wires}

\author{E.B. Magnusson}
\affiliation{Science Institute, University of Iceland, Dunhagi 3,
IS-107, Reykjavik, Iceland}

\author{I.G. Savenko}
\affiliation{Science Institute, University of Iceland, Dunhagi 3,
IS-107, Reykjavik, Iceland}
\affiliation{Academic University - Nanotechnology Research and Education Centre, 8/3 Khlopina, 194021, St.Petersburg, Russia}

\author{I.A. Shelykh}
\affiliation{Science Institute, University of Iceland, Dunhagi 3,
IS-107, Reykjavik, Iceland}
\affiliation{International Institute of Physics, Av. Odilon Gomes de Lima, 1772, Capim Macio, 59078-400, Natal, Brazil}
\date{\today}

\begin{abstract}
We investigate the phenomena of the bistability and domain wall propagation in polaritonic systems with dissipation provided by the interaction with incoherent phonon bath. The results on the temperature dependence of the polariton bistability behavior and polariton neuron switching are presented.

\end{abstract}

\maketitle

\section{Introduction}

The study of light-matter interactions is interesting and important both from the point of view of fundamental fundamental physics and device applications. In this context, the structure which attracted a particular attention is a semiconductor microcavity (MC) with a quantum well (QW) embedded into it. Tuning the energy of the excitonic transition in resonance with the energy of the photonic cavity mode one can reach a regime of strong coupling accompanied by formation of hybrid quasiparticles called exciton-polaritons. These half-light, half-matter particles exhibit a number of extraordinary properties. Due to their extremely small effective mass and bosonic nature, polaritons provide an opportunity to study various quantum collective phenomena, ranging from polariton BEC \cite{KasprzakNature} and Josephson effect \cite{LagoudakisJosephson} to polariton-mediated
superconductivity \cite{LaussySupercond}. Strong polariton-polariton interactions make it possible to observe several remarkable non-linear effects such as superfluidity \cite{AmoNature}, bi- and multi-stability \cite{bistability}, and soliton-like propagation \cite{soliton}.

\begin{figure}[!b]
\includegraphics[width=1.0\linewidth]{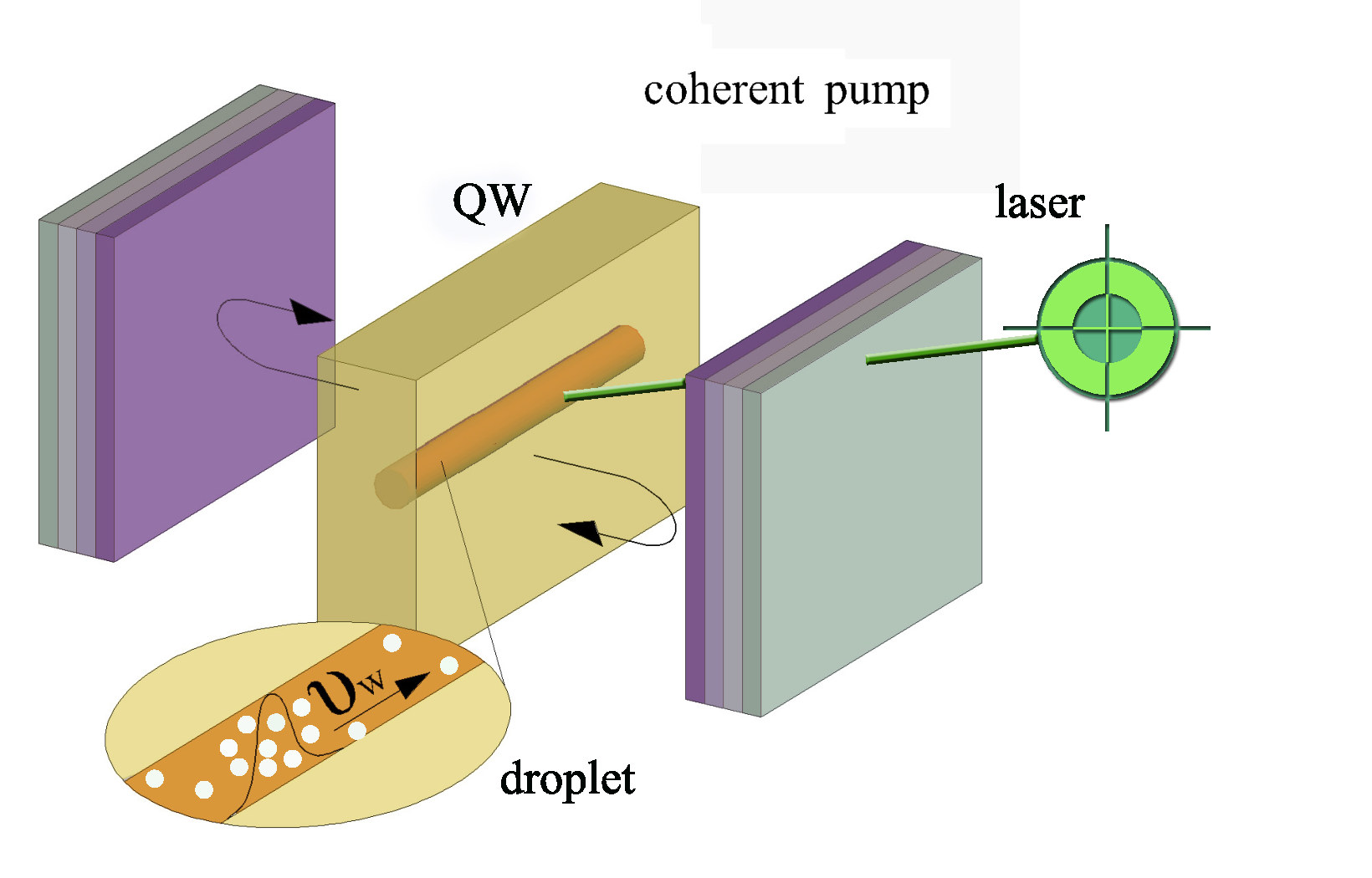}
\caption{Schematic representation of polariton wavepacket propagation along a 1D polariton wire.}
\label{Channel}
\end{figure}

Although free polaritons are 2D particles, it was recently noted \cite{WertzNature} that it can be interesting to consider 1D polariton wires (see setup in Fig.~\ref{Channel}) provided by the lateral confinement of the polaritons in one of the directions \cite{Confinement}. Such wires have the potential to become basic building blocks in future spinoptronic devices, including polariton Berry phase interferometer \cite{Berry} and polariton Datta and Das spin transistor \cite{Datta}. In the nonlinear regime, the bistability effects in 1D polariton channels allow for the realization of logical circuits based on polariton neurons \cite{LiewNeuron,LiewCircuit}, in which a local switching between states propagates throughout the wire.

The above mentioned phenomena require a proper theoretical basis for description of all relevant processes which is a non-trivial task. A successful theoretical consideration should include the effects of interaction with a reservoir of acoustic phonons which leads to polariton thermalization, and the effects of polariton-polariton scattering leading to the blueshifts and nonlinearities. Additionally, it should be taken into account that polaritons have finite lifetime, and the correct description of their dynamics should necessarily take into account the effects of pump and decay.

Currently, two main ways have been pursued to describe the dynamics of interacting polaritons. Firstly, with the assumption of full coherence, mean-field approximation gives the Gross-Pitaevskii equation (GPE) commonly used for the description of spatially inhomogeneous polariton condensates \cite{ShelykhGP}. However, while GPE includes polariton-polariton scattering it does not describe interaction with a phonon reservoir. Assuming oppositely that the polaritons are completely incoherent, the dynamics in reciprocal space can be described by the semiclassical Boltzmann equations \cite{Porras2002,Kasprzak2008,Haug2005,Cao}. Unfortunately, this technique fails to describe the real space dynamics of inhomogeneous systems.

Recently, we proposed a novel formalism based on full density matrix approach to describe the dynamics of the interacting polariton system with dissipation in real space and time \cite{Density2011}. The method was applied for the study of the propagation of 1D polariton droplets. However, we neglected the pumping terms and thus consideration of the cw regime most relevant from the experimental point of view was not done. This paper is devoted to the bridging of this evident gap. Here we investigate the phenomena of the bistability and domain wall propagation in polaritonic systems with dissipation provided by the interaction with an incoherent phonon bath and present the results on the temperature dependence of the hysteresis in the polariton system and polariton neuron switching.

\section{Formalism}

The general formalism for the time evolution of a spatially inhomogeneous bosonic system without account of the coherent pumping terms was developed in our previous paper \cite{Density2011}. We refer to this paper for details and derivations and give here only a brief overview, adding the detailed consideration of the pumping terms only.

The state of the whole system is described by the combined polariton and phonon density matrix $\rho=\rho_{ph} \otimes \rho_{pol}$ (factorization corresponds to Born approximation). The phonon part of the system is assumed to be time-independent and thermalized, $\rho_{ph}=\texttt{exp}\left\{-\beta\widehat{H}_{ph}\right\}$, while we need to determine the time-dependence of the single-particle polariton density matrix in real space. The convenient way of its representation is the use of the polariton field operators $\widehat{\psi}^\dagger(\textbf{r},t),\widehat{\psi}(\textbf{r},t)$:

\begin{eqnarray}
\rho(\textbf{r},\textbf{r}',t)=Tr\left\{\widehat{\psi}^\dagger(\textbf{r},t)\widehat{\psi}(\textbf{r}',t)\rho\right\}
\equiv\langle\widehat{\psi}^\dagger(\textbf{r},t)\widehat{\psi}(\textbf{r}',t)\rangle
\end{eqnarray}
where the trace is performed by all the degrees of freedom of the system. A Fourier transform can be performed to work in reciprocal space, which makes the calculations easier.

\begin{align}
\rho(\textbf{k},\textbf{k}',t)&=(2\pi)^d/L^d\int e^{i(\textbf{kr}-\textbf{k}'\textbf{r}')}\rho(\textbf{r},\textbf{r}',t)d\textbf{r}d\textbf{r}'=\\
\nonumber&=Tr\left\{\rho a_{\textbf{k}}^\dagger a_{\textbf{k}'}\right\}\equiv\langle a_{\textbf{k}}^\dagger a_{\textbf{k}'}\rangle
\label{matrixel}
\end{align}
where $d$ is the dimensionality of the system ($d=2$ for non-confined
polaritons, $d=1$ for the polariton channel), $L$ is its
linear size,  $a_{\textbf{k}}^\dagger$, $a_{\textbf{k}}$ are creation and
annihilation operators of the polaritons with momentum \textbf{k}. If the time dependence of the density matrix in reciprocal space is determined, an inverse Fourier transform allows to obtain the dynamics in real space straightforwardly.

The Hamiltonian of the system can be represented as a sum of the terms corresponding to various physically relevant processes in the system:
\begin{equation}
\widehat{H}=\widehat{H}_0+\widehat{H}_{pol}+\widehat{H}_{ph}+\widehat{H}_{cp}+\widehat{H}_{dcp}
\label{Hamiltonian}
\end{equation}
where
\begin{equation}
\widehat H_0=\sum_{\textbf{k}} E_\textbf{k}  a_{\textbf{k}}^\dagger a_\textbf{k}
\end{equation}
 corresponds to free polariton propagation,
 \begin{equation}
 \widehat H_{pol}=\frac{U}{2}\sum_{\textbf{k}_1,\textbf{k}_2,\textbf{p}}a_{\textbf{k}_1}^\dagger a_{\textbf{k}_2}^\dagger a_{\textbf{k}_1+\textbf{p}}a_{\textbf{k}_2-\textbf{p}}
 \end{equation}
to polariton-polariton scattering,
\begin{align}
\widehat{H}_{ph}=\sum_{\textbf{k},\textbf{q}}D(\textbf{q})a_{\textbf{k}+\textbf{q}}^\dagger a_\textbf{k}(b_\textbf{q}+b_{-\textbf{q}}^\dagger),
\end{align}
$\widehat H_{cp}$ to coherent laser pumping, $\widehat H_{dcp}$ to incoherent pumping and finite polariton lifetime (see expressions for these two terms below).

In the above formulae $E_\textbf{k}$ defines the dispersion of the free polaritons and the quantities $U$ and $D$ correspond to the polariton- polariton and polariton- phonon scattering.

The Hamiltonian \eqref{Hamiltonian} can be separated into the sum of the coherent part and the part introducing decoherence,
\begin{eqnarray}
H=H_{co}+H_{deco},\\
H_{co}=\widehat{H}_0+\widehat{H}_{pol}+\widehat{H}_{cp},\\
H_{deco}=\widehat{H}_{ph}+\widehat{H}_{dcp}
\end{eqnarray}

The effect of the coherent and incoherent parts should be treated in different ways. As for the coherent processes in the system, their effect can be accounted for using the Liouville-von Neumann equation
\begin{equation}
i\left(\partial_t\rho\right)^{(co)}=\left[\widehat H_{co};\rho\right]
\label{liouville}
\end{equation}

On the contrary, the incoherent part of the evolution is described by the Lindblad equation \cite{Carmichael} which reads

\begin{align}
\left(\partial_t\rho\right)^{(deco)}&=-\int_{-\infty}^t dt'\left[H_{deco}(t);
  \left[H_{deco}(t');\rho(t)\right]\right]= \label{Liouville_int}\\
\nonumber &= \delta_{\Delta E}\left[2\left(H^+\rho H^-+H^-\rho\right.
H^+\right)-   \nonumber \\ \nonumber
&\left.-\left(H^+H^-+H^-H^+\right)\rho-\rho\left(H^+H^-+H^-H^+\right)\right]\label{Lindblad}
\end{align}
where the coefficient $\delta_{\Delta E}$ denotes energy conservation and has dimensionality of inverse energy and is in the calculation taken to be equal to the broadening of the polariton state \cite{KavokinMalpuech}. The terms $H^+$ and $H^-$ correspond to the processes when the thermal reservoir particle (phonon or other reservoir boson, see below) is created or destroyed.

The effects of the terms corresponding to the polariton-phonon and polariton-polariton interactions was in detail considered in Ref.~\onlinecite{Density2011}. The corresponding equations for the elements of the density matrix and their derivation can be found there, and we do not reproduce them in current manuscript. For polariton- polariton scattering, these equations reproduce an analog of Gross- Pitaevskii equation written for the density matrix, and for polariton- phonon interaction they are generalizations of semiclassical Boltzmann equations.

Here we only consider the terms provided by coherent and incoherent pumps, which should be added to the dynamic equations of Ref.~\onlinecite{Density2011} to make our consideration complete.

\subsection{Coherent pumping}

We start from the case of the coherent pump. Its physical meaning is coupling to an electric field with a well defined phase, provided e.g.~by an external laser beam. Mathematically, the corresponding Hamiltonian can be introduced as
\begin{equation}
\widehat H_{cp}=\sum_{\textbf k'} p_{\textbf k'} a_{\textbf k'}^\dagger  + h.c.
\label{Hcp}
\end{equation}
The coefficients $p_\textbf k$ are Fourier transforms of the pumping amplitude in a real space by $p(\textbf x,t)$, which can be cast as
\begin{equation}
p(\textbf x,t)= P(\textbf x)e^{i\textbf k_p \textbf x} e^{-i\omega_p t}
\end{equation}
where $P(\textbf x)$ is the pumping spot profile in real space, $\textbf k_p$ is an in-plane pumping vector resulting from the inclination of the laser beam as respect to the vertical and $\omega_p$ is the pumping frequency of the single-mode laser.

Let us now check the effect of $\widehat H_{cp}$ on the evolution of the polariton density matrix. Insertion of the Hamiltonian \eqref{Hcp} into Liouville- von Neumann equation yields the following result:

\begin{align}
 \label{eq:pumping} &\partial_t \langle a^\dagger _\textbf k a_\textbf k \rangle = -\frac{2}{\hbar} \textrm{Im} \{ p_\textbf k^* \langle a_\textbf k \rangle \}  \\ \nonumber
 &\partial \langle a_{\textbf k_1}^\dagger  a_{\textbf k_2} \rangle= \frac{i}{\hbar} ( p_{\textbf k_1}^* \langle a_{\textbf k_2} \rangle - p_{\textbf k_2} \langle a_{\textbf k_1}  \rangle^* )
\end{align}

One sees, that the equations for matrix elements contain a new quantity- an average value of the annihilation operator of the polariton field, which is nothing but the order parameter (also called a macroscopic wavefunction) of the system. One thus needs to obtain the dexpression for this quantity to close the system of the equations. Straightforward derivation gives:

\begin{align}
 \label{eq:pumping1}
& \partial_t \langle a_\textbf k \rangle = -\frac{i}{\hbar} p_\textbf k  -\frac{i}{\hbar} E_\textbf k \langle a_\textbf k \rangle  \\ \nonumber
 &-\frac{i}{\hbar} U \sum_{\textbf k_2,\textbf p} \rho(\textbf k_2,\textbf k_2-\textbf p) \langle a_{\textbf k+\textbf p} \rangle \\ \nonumber
 &+\left(\sum_{\textbf q,E_\textbf k<E_{\textbf k+\textbf q}}  W(\textbf q)   (\rho(\textbf k+\textbf q,\textbf k+\textbf q) -  n_\textbf q^{ph}) \right.\\ \nonumber
&+\left. \sum_{\textbf q,E_\textbf k>E_{\textbf k+\textbf q}}  W(\textbf q)  (-\rho(\textbf k+\textbf q,\textbf k+\textbf q) -  n_\textbf q^{ph}-1)\right)\langle a_\textbf k \rangle
\end{align}
where $U$ is a matrix element of polariton- polairton interaction and $W(\textbf q)$ are transition rates of phonon- assisted processes (See \cite{Density2011} for the definition of the parameters).

\subsection{Incoherent pumping and lifetime}

By incoherent pump we mean the exchange of particles between the polariton system and some incoherent bosonic reservoir, whose nature depends on the pumping scheme. Usually, this will be an ensemble of incoherent excitons created either by an electrical pump or by incoherent optical excitation or a reservoir of the external photonic modes providing the leakage of the photons from the cavity. The corresponding Hamiltonian written in Dirac representation reads

\begin{eqnarray}
H_{dcp}=\sum_{\textbf{k},\textbf{k}'}K(\textbf{k},\textbf{k}')e^{i(E_\textbf{k}-E_{R,k})t/\hbar}a_\textbf{k}^\dagger b_{\textbf{k}'}+H.c.=\\
\nonumber=H_{icp}^++H_{icp}^-
\label{Hdcp}
\end{eqnarray}
where $b_\textbf{k}$ is a secondary quantization operator corresponding to the bosonic reservoir in question, $K(\textbf{k},\textbf{k}')$ are constants characterizing the coupling between the polariton system and the reservoir  The introduction of this Hamiltonian into Lindblad equation leads to the standard terms, whose derivation can be found elsewhere\cite{Carmichael,Laussy2008}:

\begin{equation}
\label{eq:incoherent} \partial_t \langle a^\dagger_{\textbf{k}} a_{\textbf{k}'} \rangle= I_{\textbf k}\delta_{\textbf k \textbf k '}- \frac{1}{2\hbar}(\gamma_\textbf k+\gamma_{\textbf k '})\langle a^\dagger_\textbf k a_{\textbf k '} \rangle
\end{equation}
where the terms $I_\textbf k$ and $\gamma_\textbf k$ denote the intensity of the incoherent pump and broadening of the polaritonic levels connected with lifetimes of the polariton states, $\gamma_\textbf k=\hbar\tau_\textbf k^{-1}$. They are usually taken as phenomenological parameters, but can be connected with the quantities entering in Hamiltonian \eqref{Hdcp}:
\begin{eqnarray}
I_\textbf k=\frac{1}{\hbar}\sum_{\textbf k'}|K(\textbf{k},\textbf{k}')|^2\delta(E(\textbf{k})-E'(\textbf{k}'))n_{\textbf{k}'},\\
\gamma_\textbf k=\sum_{\textbf k'}|K(\textbf{k},\textbf{k}')|^2\delta(E(\textbf{k})-E'(\textbf{k}'))
\end{eqnarray}
where $n_{\textbf{k}'}$ are the occupancies of the bosonic reservoir.

In our further consideration we will consider only the case of the coherent pump, thus putting all $I_{\textbf{k}}=0$ and retaining in the resulting equations only the terms corresponding to the lifetime, which corresponds to the case of the empty bosonic reservoir , $n_{\textbf{k}}=0$ for all the \textbf k.

\section{Results and discussion}

To get a full system of equations for the dynamics of the polariton system with pump and decay, one should combine the equations derived in Ref.~\onlinecite{Density2011} with expressions \eqref{eq:pumping},\eqref{eq:pumping1},\eqref{eq:incoherent}. The resulting formalism is suitable both for describing 2D polaritons and polaritons confined in 1D channels. The consideration of the former case, however, needs powerful computing facilities and in the present paper we focus only on consideration of the latter one.

We considered a $2$ $\mu m$ wide polariton channel in a microcavity with an active region based on InAlGaAs alloys with Rabi splitting 15 meV. The matrix elements of the polariton- polariton and polariton- phonon interactions were estimated using the standard formulae \cite{Ciuti,Golub}.

The first phenomenon we modeled was the effect of the thermalization in the polariton system provided by polariton- phonon interaction on its bistable behavior. It is well known that if the polariton ensemble is fully coherent and its dynamics is described by the Gross- Pitaevskii equation containing coherent pumping and lifetime terms, and if the energy of the pumping laser lies slightly above the bottom of the lower polariton branch, the dependence of the concentration of the polaritons on pump intensity is described by an S-shaped curve, characteristic for systems revealing the effects of bistability and hysteresis \cite{Gippius}. Such a behavior is due to the polariton- polariton interactions which introduce nonlinearity into the system. On the other hand, in the approach based on the semiclassical Boltzmann equations corresponding to the limit of strong decoherence the bistability is absent and the dependence of the occupancy of the ground state on the pump intensity is described by a single defined threshold function \cite{Kasprzak2008}. One can expect that a transition between these two regimes should occur if one raises the temperature in the system which leads to the intensification of the polariton- phonon interactions and decoherence in the system. This was indeed observed in our calculations.

\begin{figure}
\includegraphics[width=0.85\linewidth]{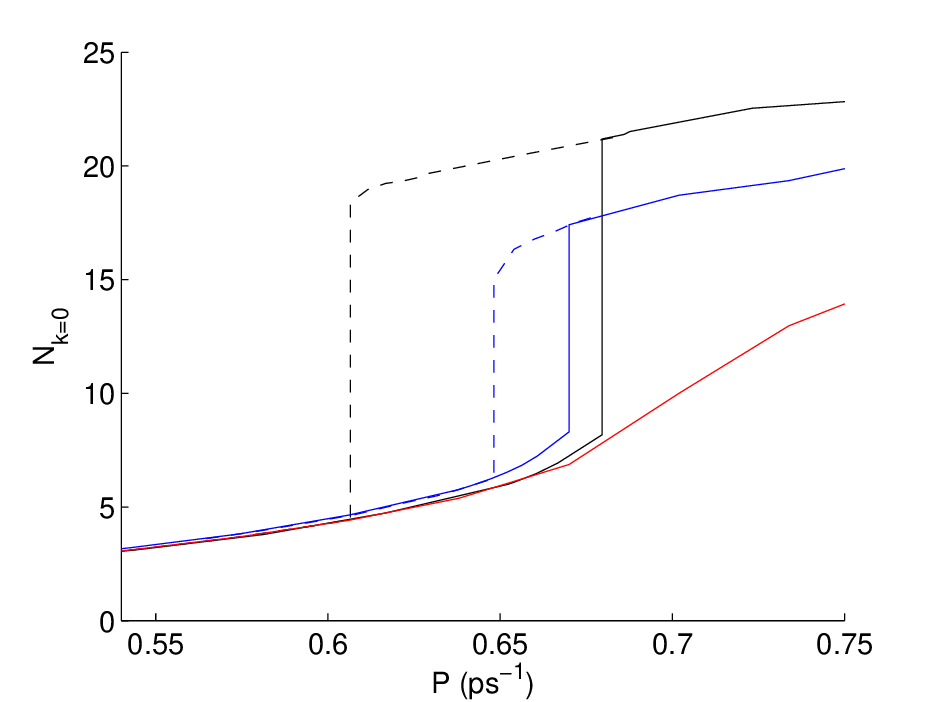}
\caption{
The dependence of the polariton concentration on spatially homogeneous cw pumping intensity for various temperatures. At 20K, the hysteresis curve is quite wide (black). At 30K, the hysteresis curve is much narrower, while at 50K it has disappeared completely.
} \label{fig:bistability}
\end{figure}

The computational results are presented in Fig.~\ref{fig:bistability}. The system is pumped with a spatially homogenious laser beam oriented  perpendicular to the QW (i.e. at $k=0$ in $k$-space) at a slightly higher frequency than the $k=0$ polariton frequency (0.24 meV detuning). At low pump intensity, the pump frequency is not in resonance with the condensate. Consequently, the condensate occupation remains fairly low. As the pumping is increased, the polariton energy is blueshifted into resonance with the pump and there is a sudden jump in the occupation of the $k=0$ polariton state at some characteristic pump intensity $I_0$. If one then decreases the pump, the polariton occupancy jumps down at different value $I_1<I_0$ which corresponds to the hysteresis behavior. However, the increase of the temperature leads to the intensification of the phonon scattering, and bistable behavior becomes less and less pronounced: the hysteresis area narrows and is quenched completely above some critical temperature $T_c\approx40K$. This correspond to the transition between the Gross-Pitaevskii and Boltzmann regimes in the polariton system. A similar phenomenon was earlier predicted for microcavity- based terahertz emitting device \cite{Terahertz}.

The effect of bistability can form a basis for creation of a variety of devices based on nonlinear polariton transport \cite{PolaritonDevices}. Among them are polariton neurons \cite{LiewNeuron}- the building blocks of polariton based optical integrated circuits \cite{LiewCircuit} utilizing the phenomenon of domain wall propagation in bistable systems. The underlying idea is following. Imagine that the polariton system is driven by a spatially homogeneous background cw pump with intensity corresponding to the bistable regime and the steady state occupancy of the system corresponds to the lower branch of the S- shaped curve. Then a short localized pulse is applied in the middle of the wire. Its intensity should be enough to send the condensate locally to the upper branch of the S-shaped curve. Due to diffusion, the polariton wavepacket spreads to the neighboring regions and switches them to the upper branch. This way, the area of high occupancy steadily expands. This phenomenon is analogous to the propagation of the domain wall in ferromagnetic materials. It should be noted that although the polaritons have finite lifetime, this does not limit the length of signal propagation in a polariton neuron, and the signal keeps propagating as long as the background cw pumping persists.

\begin{figure}
\includegraphics[width=0.8\linewidth]{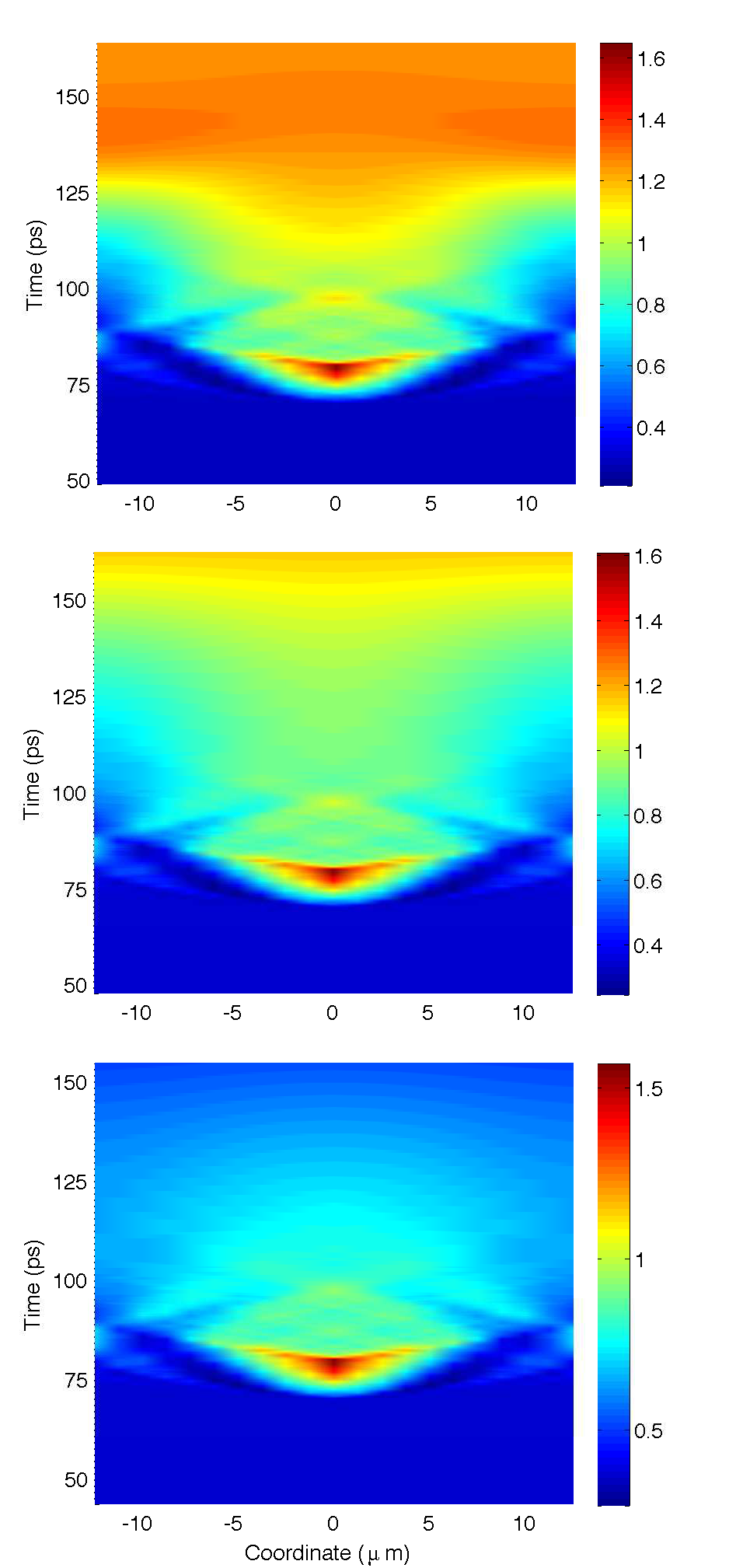} 
\caption{
Neuron behavior and domain wall propagation for (top to bottom) T=15K, 20K and 25K. The plots show the polariton concentration in real space (horizontal axis) and time (vertical axis). The system is pumped non-resonantly by spatially homogenious cw laser, at 70-80ps a switching pulse arrives. At 15 K one sees the propagation of the distinctive domain wall which becomes more smeared and fully disappears at 25 K due to that there is no bistability at this pump intensity and temperature.
} \label{fig:neuron}
\end{figure}

As bistability in the polaritonic system strongly depends on temperature as was discussed above, the same should be true for the domain wall propagation in the polariton neurons. This can indeed be seen from Fig.~\ref{fig:neuron} representing the results of our calculations.
At 15K, the switching is clear and one can easily detect the propagation of a sharp well defined domain wall. At 20K and using the same pump intensity as before, the switching is less pronounced, partly because the higher intensity is lowered and partly because the spreading of the high intensity is slower, which leads to the washing out of the domain wall. At 25K there is no bistability for this pump intensity anymore, and the system goes back to its low population state after the application of the pulse.

\begin{figure}
\includegraphics[width=0.8\linewidth]{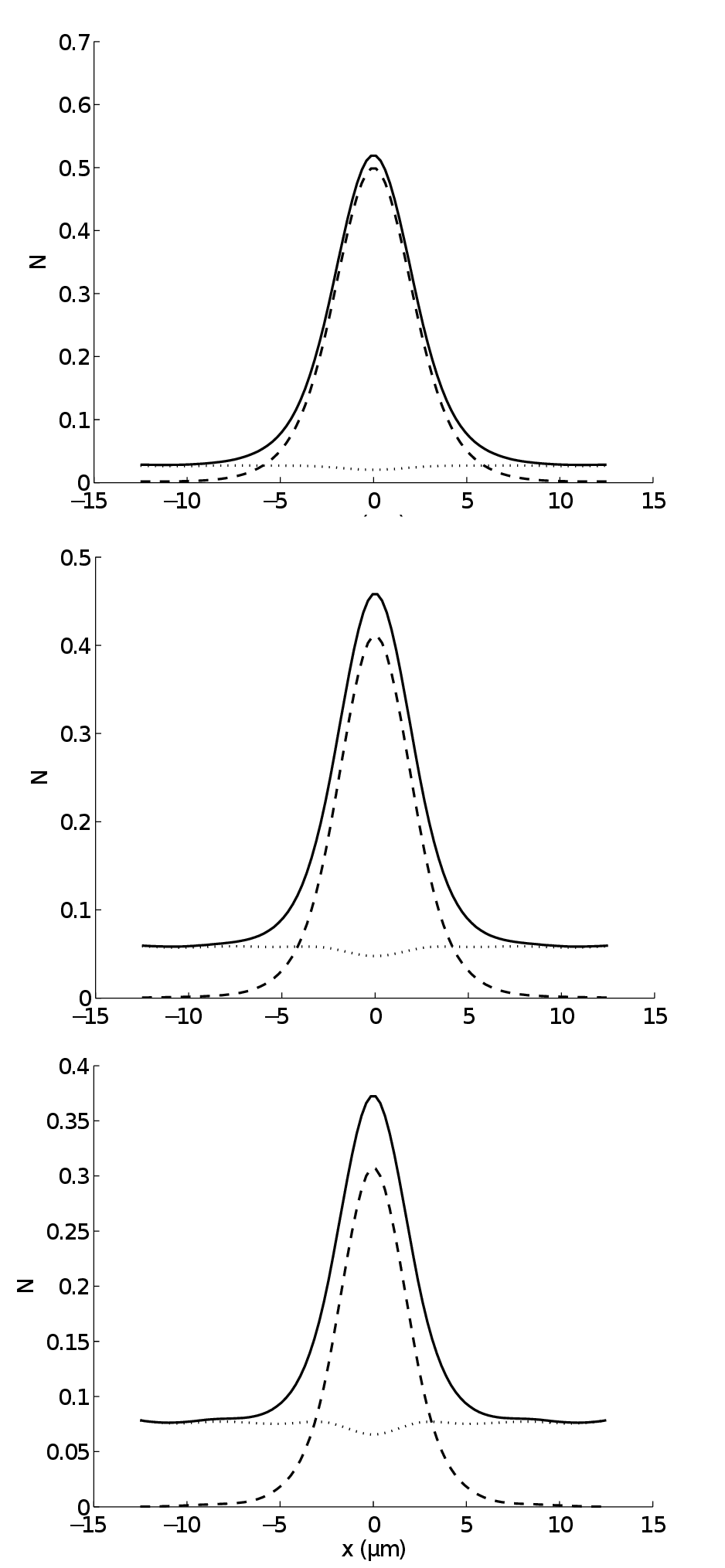}
\caption{
Coherent fraction at (top to bottom) T=10K, 30K and 60K. Pumping at $k=0$, no detuning. The solid lines show the total population, the dashed lines the coherent population and the dotted ones the incoherent population.
} \label{fig:decoherence}
\end{figure}

Finally, the effect of phonon scattering on pure dephasing in the system was considered. We pumped the system with a coherent pulse having Gaussian profile in the real space. For various temperatures of the system state we investigated in the steady state the spatial profiles of both total polariton density and its coherent part determined as $| \psi (x) |^2$, where
$$
\psi(x)= \int_{-\infty}^{+\infty} \langle a_k \rangle e^{ikx}dk
$$
The results are shown in Fig. \ref{fig:decoherence}. One can see that at low temperatures the coherent fraction is quite large at the center of the coherent pumping spot, but it dramatically decays outside it. However, as temperature is increased, the density profile gets more spread out over the wire, forming an almost constant background density made up of the decoherent population.

\section{Conclusion}
In conclusion, we have considered the effects of coherent pumping and finite lifetime in a polaritonic system accounting for all physically relevant processes in the system. We applied our theory for the consideration of nonlinear polariton propagation in a 1D polariton wire. We have shown that the increase of temperature dramatically affects such processes as bistability switching and domain wall propagation in polariton neurons.

\section{Acknowledgements}

We thank Dr. T.C.H. Liew, Dr. G. Malpuech, Dr. D.D. Solnyshkov and Mr. O. Kyriienko for useful discussions. The work was supported by Rannis "Center of excellence in polaritonics" and FP7 IRSES project "POLAPHEN". I.A.S acknowledges the support from COST POLATOM project and thanks Mediterranean Institute of Fundamental Physics for hospitality.

\end{document}